\documentclass[conference]{IEEEtran}
\IEEEoverridecommandlockouts
\usepackage{graphicx}
\usepackage{epstopdf} %converting to PDF
\usepackage{verbatim} %+程式碼
\usepackage{multirow} %+表格
\usepackage{subfigure}
\usepackage{colortbl} %灰階表格
\usepackage{indentfirst}
\usepackage{dcolumn,booktabs}
\usepackage{algorithm}
\usepackage[noend]{algpseudocode}

\usepackage{algorithmicx,algorithm}

\usepackage{algpseudocode}
\usepackage{graphics}
\usepackage{epsfig,color}
\usepackage{amsmath}
\usepackage{amssymb}
\ifCLASSINFOpdf

\else
\fi
\newtheorem{Proposition}{Proposition}

\begin{document}

\title{Generalized Expectation Consistent Signal Recovery for Nonlinear Measurements}

%\author{
%\IEEEauthorblockN{Hengtao He\IEEEauthorrefmark{1},  Chao-Kai Wen\IEEEauthorrefmark{2}, Shi Jin\IEEEauthorrefmark{1}}
%\IEEEauthorblockA{\IEEEauthorrefmark{1}National Mobile Communications Research Laboratory, Southeast University\\ Nanjing 210096, P. R. China, E-mail: \{hehengtao, jinshi\}@seu.edu.cn}
%\IEEEauthorblockA{\IEEEauthorrefmark{2}Institute of Communications Engineering, National Sun Yat-sen University\\ Kaohsiung 804, Taiwan, E-mail: chaokai.wen@mail.nsysu.edu.tw}
%}
\author{Hengtao He, Chao-Kai Wen, Shi Jin\thanks{Hengtao He and S. Jin are with the National Mobile Communications Research
Laboratory, Southeast University, Nanjing 210096, China (e-mail: hehengtao@seu.edu.cn, and jinshi@seu.edu.cn).}
\thanks{C.-K. Wen is with the Institute of Communications Engineering, National
Sun Yat-sen University, Kaohsiung 804, Taiwan (e-mail: chaokai.wen@mail.nsysu.edu.tw).}
\thanks{The work of Hengtao He and S. Jin was supported in part by the National Science Foundation (NSFC) for Distinguished Young Scholars of China with Grant 61625106 and the National Natural Science Foundation of China under Grant 61531011. The work of C.-K. Wen was supported by the ITRI in Hsinchu, Taiwan, and the MOST of Taiwan under Grants MOST 103-2221-E-110-029-MY3.}
}

\maketitle

\begin{abstract}

In this paper, we propose a generalized expectation consistent signal recovery algorithm to estimate the signal $\mathbf{x}$ from the nonlinear measurements of a linear transform output $\mathbf{z}=\mathbf{A}\mathbf{x}$.
This estimation problem has been encountered in many applications, such as communications with front-end impairments, compressed sensing, and phase retrieval.
The proposed algorithm extends the prior art called generalized turbo signal recovery from a partial discrete Fourier transform matrix $\mathbf{A}$ to a class of general matrices.
Numerical results show the excellent agreement of the proposed
algorithm with the theoretical Bayesian-optimal estimator derived using the replica method.
\end{abstract}


% \fi
\IEEEpeerreviewmaketitle

\section{Introduction}
Signal reconstruction problems are encountered in many engineering fields.
Compressed sensing (CS) \cite{Donoho_CS,intrudection_compressed_sensing} aims to
reconstruct a sparse signal with a high-dimension space from a low-dimension measurement space. Significant attention has been given to the usage of $l_{1}$-norm minimization because it is capable of recovering sparse signal with a computational cost of polynomial complexity. However, this approach is still generally far from optimal \cite{lp_norm}.

Given that the prior distribution of the signal is used, the Bayesian inference offers an optimal recovery approach in the minimum mean square error (MMSE) perspective although its exact execution is computationally difficult in most cases \cite{Bayesian_Compressive_Sensing}. % Along with this line, sparse Bayesian learning was derived from research area of machine learning and has become a popular method for sparse signal recovery in CS.
Approximate message passing (AMP), which is based on the Gaussian approximations of loopy belief propagation, is a tractable and less complex alternative, and it has attracted considerable attention for such problems \cite{Donohol_AMP,Krzakala_AMP}. Unfortunately, AMP and its generalization, GAMP {\cite{SRangan_GAMP}}, are fragile in terms of the choice of matrix, and can perform poorly outside the special case of zero-mean, i.i.d., sub-Gaussian matrix. %{\cite{Adaptive_damping_GAMP}}.

Ma et al. \cite{JMa} developed a signal recovery (SR) algorithm under \emph{linear} measurements called Turbo-SR with partial discrete Fourier transform (DFT) as the sensing matrix. Subsequently Liu et al. \cite{LiuT} proposed the generalized Turbo-SR (GTurbo-SR) to address \emph{non-linear} measurements. Ma and Li \cite{OAMP} further proposed the orthogonal AMP (OAMP) algorithm for general sensing matrices but under linear measurements.
In contrast to suboptimal developments along this line, such as AMP and GAMP, Turbo-SR, GTurbo-SR, and OAMP are optimal and have excellent convergence properties.
The state evolutions of the three algorithms agree perfectly with those predicted by the theoretical replica method.
However, these algorithms only consider either the partial DFT sensing matrix or linear measurements.

The purpose of this paper is to develop a novel algorithm for Bayesian SR with a much broader class of sensing matrices under non-linear measurements. We employ an advanced mean field method known as the expectation consistent (EC) approximation developed in statistical mechanics \cite{Opper_TAP,Opper_EC} and machine learning \cite{EP}. Recently, ``vector AMP" which is presented in \cite{VAMP}, can be interpreted as an instance of the generalized EC (GEC) \cite{GEC} algorithm.

Our wok is inspired by \cite{GEC}. Specifically, we present the GEC-SR to recover sparse signals from nonlinear measurements, especially from low-resolution quantized output, which has been of particular interest in recent years. We show that the performance of our GEC-SR is superior to ``initial GEC" \cite{GEC} because of different update manner.\footnote {One can introduce various iterative algorithms to the EC approximation. However, a proper update manner is important because an improper one might result in a poor convergence in particular for small measurement ratio.} When partial DFT matrix is considered, the GEC-SR is reduced to GTurbo-SR \cite{LiuT}. In addition, we give the state evolution (SE) analysis and show that the analytical SE of the GEC-SR is consistent with that obtained by the replica method. This consistency
indicates the optimality of the GEC-SR for non-linear measurements with general sensing matrices.

\emph{Notations}---For any matrix $\mathbf{A}$, $\mathbf{A}^{H}$ is the conjugate transpose of $\mathbf{A}$, and ${\sf tr}(\mathbf{A})$ denotes the traces of $\mathbf{A}$. In addition, $\mathbf{I}$ is the identity matrix, $\mathbf{0}$ is the zero matrix,
$\mathrm{Diag}(\mathbf{v})$ is the diagonal matrix whose diagonal equals $\mathbf{v}$, $\mathbf{1}_n$ is the $n$-dimensional all-ones vector, $\mathbf{d}(\mathbf{Q})$ is the diagonalization operator, which returns a constant vector containing the average diagonal elements of $\mathbf{Q}$, and $<\mathbf{a}>$ is the average operator, which returns a constant vector containing the average elements of $\mathbf{a}$. In addition, $\oslash$ and $\odot$ denote componentwise vector division and vector multiplication, respectively.
%and $\mathtt{E}\{\cdot\}$, $\mathtt{Var}\{\cdot\}$ represent the expectation, and variance operators, respectively.
 A random vector $\mathbf{z}$ drawn from the proper complex Gaussian distribution of mean $\boldsymbol{\mu}$ and covariance $\boldsymbol{\Omega}$ is described by the probability density function:
\begin{equation*}
  \mathcal{N}_{\mathbb{C}}(\mathbf{z};\boldsymbol{\mu},\boldsymbol{\Omega})=\frac{1}{\mathrm{det}(\pi \boldsymbol{\Omega})}
  e^{-(\mathbf{z}-\boldsymbol{\mu})^{H}\boldsymbol{\Omega}^{-1}(\mathbf{z}-\boldsymbol{\mu})}.
\end{equation*}
We use $D z$ to denote the real Gaussian integration measure
\begin{equation*}
  Dz=\phi(z)dz, \quad \phi(z)\triangleq\frac{1}{\sqrt{2\pi}}e^{-\frac{z^{2}}{2}},
\end{equation*}
and we use $Dz_{c}=\frac{e^{-|z|^{2}}}{\pi}dz$ to denote the complex Gaussian integration measure. Finally, $\Phi(x)\triangleq \int_{-\infty}^{x} Dz$
denotes the cumulative Gaussian distribution function.

\section{Problem Description}
\subsection{Observation Model}
We consider the generalized linear model (GLM) where a $N$-dimensional random vector $\mathbf{x}\in\mathbb{C}^{N} $ is observed through a linear output $\mathbf{z}=\mathbf{A}\mathbf{x}$, followed by a componentwise, probabilistic measurement channel
\begin{equation}\label{eq1}
  p(\mathbf{y}|\mathbf{x})=\prod\limits_{m=1}^{M}p(y_{m}|z_{m}),\quad \mathbf{z}=\mathbf{A}\mathbf{x},
\end{equation}
where $\mathbf{A}\in \mathbb{C}^{M \times N}$ is a known transform matrix. The sparse signal $\mathbf{x}$ is assumed to be i.i.d. with the $n$th entry of $\mathbf{x}$ following the Bernoulli-Gaussian distribution:
\begin{equation}\label{eq2}
  p(x)=(1-\rho)\delta(x)+\rho\mathcal{N}_{\mathbb{C}}(x;0,\rho^{-1}),
  \end{equation}
where $\delta(x)$ is the Dirac function, and the variance of each $x_{n}$ is normalized, that is, $ \mathtt{E}\{|x_{n}|^{2}\}=1$. We denote the measurement ratio by $\alpha = M/N$ (i.e., the number of measurements per variable). In addition, for ease of notation, we define
\begin{equation}\label{eq:defPxPz}
  P_{x} = \mathtt{E}\{|x_{n}|^{2}\} ~~\mbox{and}~~  P_{z} = P_{x} \cdot {\sf tr}(\mathbf{A}\mathbf{A}^H)/M .
\end{equation}
\subsection{Quantized Measurements}
In this study, we are interested in the measurements acquired through the complex-valued quantizer $Q_{c}$. Specifically, each complex-valued quantizer $Q_{c}$ consists of two real-valued $B$-bit quantizers $Q$, which is defined as
\begin{equation}\label{eq3}
  \tilde{y}_{m}=Q_{c}(y_{m})\triangleq Q(y_{R,m})+jQ(y_{I,m}).
\end{equation}
Therefore, the resulting quantized signal $\tilde{\mathbf{y}}$ is provided by
\begin{equation}\label{eq4}
  \tilde{\mathbf{y}}=Q_{c}(\mathbf{\mathbf{y}})=Q_{c}(\mathbf{z}+\mathbf{w}),
\end{equation}
where $\mathbf{w}\sim\mathcal{N}_{\mathbb{C}}(\mathbf{0},\sigma^{2}\mathbf{I})$ represents the additive Gaussian noise.
The output is assigned the value $\tilde{y}_{m}$ when the quantizer input falls in the corresponding interval $(\tilde{y}_{m}^{\rm low},\tilde{y}_{m}^{\rm up}]$ (namely, the $b$-th bin). For example, the quantized output of a typical uniform quantizer with a quantizer step size $\Delta$ is given by
\begin{equation}\label{eq5}
  \tilde{y}_{m}\in{\left\{{\left(-\frac{1}{2}+b\right)}\Delta; \, b=-\frac{2^{B}}{2}+1,\cdots,\frac{2^{B}}{2}\right\}}\triangleq \mathcal{R}_{B},
\end{equation}
and the associated lower and upper thresholds are given by
\begin{align}
\tilde{y}_{m}^{\rm low} &=
  \begin{cases}
   \tilde{y}_{m}-\frac{\Delta}{2} , & \mbox{if } \tilde{y}_{m}\geq-{\left(\frac{2^{B}}{2}-1\right)}\Delta, \\
    -\infty, & \mbox{otherwise}.
\end{cases}\label{eq6} \\
\tilde{y}_{m}^{\rm up} &=
  \begin{cases}
   \tilde{y}_{m}+\frac{\Delta}{2} , & \mbox{if } \tilde{y}_{m}\leq {\left(\frac{2^{B}}{2}-1\right)}\Delta, \\
    \infty, & \mbox{otherwise}.
\end{cases} \label{eq7}
\end{align}

We suppose that each entry of $\mathbf{x}$ is generated from a distribution (2) independently, that is,  $p(\mathbf{x})=\prod\limits_{n=1}^{N}p(x_{n})$.
%\begin{equation}\label{eq8}
%  p(\mathbf{x})=\prod_{n=1}^{N}p(x_{n}).
%\end{equation}
The componentwise, probabilistic measurement channel is given by
%\begin{equation}\label{eq9}
%  p(\tilde{\mathbf{y}}|\mathbf{z})=\prod\limits_{m=1}^{M}p(\tilde{y}_{m}|z_{m}),
%\end{equation}
%where
\begin{equation}\label{eq10}
p(\tilde{y}_{m}|z_{m})=\Psi{\left(\tilde{y}_{R,m};z_{R,m},\frac{\sigma^{2}}{2}\right)}\Psi{\left(\tilde{y}_{I,m};z_{I,m},\frac{\sigma^{2}}{2}\right)},\\
\end{equation}
where
\begin{equation}\label{eq11}
\Psi(\tilde{y};z,c^2)=\Phi{\left(\frac{\tilde{y}^{\rm up}-z}{ c}\right)}-\Phi{\left(\frac{\tilde{y}^{\rm low}-z}{ c}\right)}.
\end{equation}

\begin{figure*}
  \centering
  \includegraphics[width=15cm]{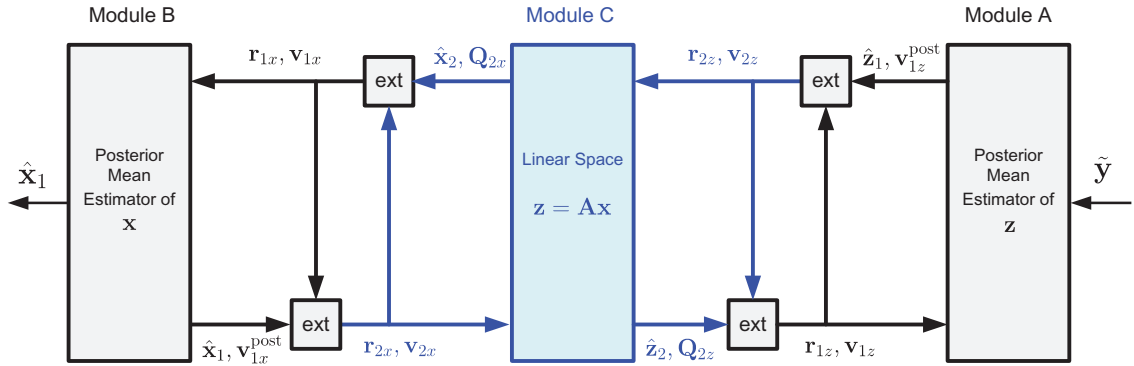}
  \caption{Block diagram of the GEC-SR algorithm}\label{GEC-diagram}
\end{figure*}

\section{Generalized EC Signal Recovery}
In this section, we present the GEC-SR. The block diagram of the GEC-SR is illustrated in Figure \ref{GEC-diagram}, which consists of three modules: modules A, B and C. Module A
computes the posterior mean and variance of $\mathbf{z}$, module C constrains the estimation into the linear space $\mathbf{z} = \mathbf{A}\mathbf{x}$, and module B computes the posterior mean and variance of $\mathbf{x}$.
These procedures follow a circular manner, that is, $A \rightarrow C \rightarrow B \rightarrow C \rightarrow A \rightarrow \cdots$.
In addition, each module uses the turbo principle in iterative decoding, that is, each module passes the extrinsic messages to its next module. The GEC-SR is different from
the GTurbo-SR \cite{LiuT} and ``initial GEC'' \cite{GEC}. We will discuss their differences in the following subsections.

\begin{algorithm}%[t]
\begin{footnotesize}
\caption{GEC-SR for the GLM} %算法的名字
\hspace*{0.02in} {\bf Input:} %算法的输入， \hspace*{0.02in}用来控制位置，同时利用 \\ 进行换行
Nonlinear measurements $\tilde{\mathbf{y}}$, sensing matrix $\mathbf{A}$, likelihood $p(\tilde{\mathbf{y}}|\mathbf{z})$, and prior distribution $p(\mathbf{x})$.\\
\hspace*{0.02in} {\bf Output:} %算法的结果输出
Recovered signal $\hat{\mathbf{x}}_{1}$.\\
\hspace*{0.02in} {\bf Initialize:} %算法的结果输出
$t \leftarrow 1$, $\mathbf{r}_{1\mathbf{z}}\leftarrow \mathbf{0}$, $\mathbf{r}_{2\mathbf{x}}\leftarrow \mathbf{0}$, $\mathbf{v}_{1\mathbf{z}}\leftarrow P_{z} \mathbf{1}$, and $\mathbf{v}_{2\mathbf{x}}\leftarrow P_{x} \mathbf{1}$.
\begin{algorithmic}[1]
\While{$t<T_{\max}$}

 \begin{enumerate}
 \item Compute the posterior mean and covariance of $\mathbf{z}$% Compute the posterior mean and covariance of $\mathbf{x}$ and $\mathbf{z}$
   \begin{subequations} \label{eq27}
   \begin{align}
     \hat{ \mathbf{z}}_{1} &=\mathtt{E}\left\{\mathbf{z}|\mathbf{\mathbf{r}}_{1\mathbf{z}},\mathbf{v}_{1\mathbf{z}} \right\}, \label{eq27a} \\
     \mathbf{v}_{1\mathbf{z}}^{\rm post} &= \mathtt{Var} \left\{ \mathbf{z}|\mathbf{\mathbf{r}}_{1\mathbf{z}},\mathbf{v}_{1\mathbf{z}} \right\}. \label{eq27b}
  \end{align}
  \end{subequations}
 Compute the extrinsic information of $\mathbf{z}$% Compute the posterior mean and covariance of $\mathbf{x}$ and $\mathbf{z}$
    \begin{subequations}\label{eq28}
    \begin{align}
     % \boldsymbol{\eta}_{1\mathbf{z}}&=\mathbf{1} \oslash \mathbf{d}(\mathbf{Q}_{1\mathbf{z}}), \label{eq28a} \\
     \mathbf{v}_{2\mathbf{z}}&= \mathbf{1}\oslash {\left( \mathbf{1}\oslash<\mathbf{v}_{1\mathbf{z}}^{\rm post}>- \mathbf{1}\oslash\mathbf{v}_{1\mathbf{z}} \right)}, \label{eq28b} \\
     \mathbf{r}_{2\mathbf{z}}&= \mathbf{v}_{2\mathbf{z}} \odot {\left( \hat{\mathbf{z}}_{1}\oslash<\mathbf{v}_{1\mathbf{z}}^{\rm post}>-\mathbf{r}_{1\mathbf{z}}\oslash\mathbf{v}_{1\mathbf{z}} \right)}. \label{eq28c}
    \end{align}
  \end{subequations}
 \item Compute the mean and covariance of $\mathbf{x}$  from the linear space
 \begin{subequations}\label{eq29}
    \begin{align}
     \mathbf{Q}_{2\mathbf{x}}&={\left(\mathrm{Diag}(\mathbf{1}\oslash \mathbf{v}_{2\mathbf{x}})+\mathbf{A}^{H}\mathrm{Diag}(\mathbf{1}\oslash\mathbf{v}_{2\mathbf{z}})\mathbf{A} \right)}^{-1}, \label{eq29a}\\
     \hat{\mathbf{x}}_{2}&=\mathbf{Q}_{2\mathbf{x}} \left(\mathbf{r}_{2\mathbf{x}}\oslash\mathbf{v}_{2\mathbf{x}}+\mathbf{A}^{H}\mathbf{r}_{2\mathbf{z}}\oslash\mathbf{v}_{2\mathbf{z}}\right). \label{eq29b}
 \end{align}
\end{subequations}
Compute the extrinsic information of $\mathbf{x}$%from the linear space
  \begin{subequations}\label{eq30}
    \begin{align}
     % \boldsymbol{\eta}_{2\mathbf{x}}&=\mathbf{1}\oslash\mathbf{d}(\mathbf{Q}_{2\mathbf{x}}), \label{eq30a} \\
     \mathbf{v}_{1\mathbf{x}}&= \mathbf{1}\oslash \left( \mathbf{1}\oslash\mathbf{d}(\mathbf{Q}_{2\mathbf{x}}) - \mathbf{1}\oslash\mathbf{v}_{2\mathbf{x}} \right), \label{eq30b} \\
     \mathbf{r}_{1\mathbf{x}}&=\mathbf{v}_{1\mathbf{x}} \odot \left( \hat{\mathbf{x}}_{2}\oslash\mathbf{d}(\mathbf{Q}_{2\mathbf{x}}) - \mathbf{r}_{2\mathbf{x}}\oslash\mathbf{v}_{2\mathbf{x}} \right). \label{eq30c}
    \end{align}
\end{subequations}
  \item Compute the mean and covariance of $\mathbf{x}$% Compute the posterior mean and covariance of $\mathbf{x}$ and $\mathbf{z}$
  \begin{subequations}\label{eq31}
      \begin{align}
      \hat{\mathbf{x}}_{1}&=\mathtt{E}\left\{\mathbf{x}|\mathbf{\mathbf{r}}_{1\mathbf{x}},\mathbf{v}_{1\mathbf{x}}\right\}, \label{eq31a} \\
      \mathbf{v}_{1\mathbf{x}}^{\rm post}&=\mathtt{Var} \left\{ \mathbf{x}|\mathbf{\mathbf{r}}_{1\mathbf{x}},\mathbf{v}_{1\mathbf{x}} \right\}. \label{eq31b}
      \end{align}
  \end{subequations}
 Compute the extrinsic information of $\mathbf{x}$
  \begin{subequations}\label{eq32}
    \begin{align}
     % \boldsymbol{\eta}_{1\mathbf{x}}&=\mathbf{1}\oslash\mathbf{d}(\mathbf{Q}_{1\mathbf{x}}), \label{eq32a} \\
     \mathbf{v}_{2\mathbf{x}}&= \mathbf{1}\oslash {\left( \mathbf{1}\oslash<\mathbf{v}_{1\mathbf{x}}^{\rm post}>- \mathbf{1}\oslash\mathbf{v}_{1\mathbf{x}} \right)}, \label{eq32b} \\
     \mathbf{r}_{2\mathbf{x}}&=\mathbf{v}_{2\mathbf{x}} \odot {\left( \hat{\mathbf{x}}_{1}\oslash<\mathbf{v}_{1\mathbf{x}}^{\rm post}> - \mathbf{r}_{1\mathbf{x}}\oslash\mathbf{v}_{1\mathbf{x}} \right)} . \label{eq32c}
    \end{align}
    \end{subequations}
\item Compute the mean and covariance of $\mathbf{z}$ from the linear space
 \begin{subequations}\label{eq33}
    \begin{align}
     \mathbf{Q}_{2\mathbf{x}}&={\left(\mathrm{Diag}(\mathbf{1}\oslash \mathbf{v}_{2\mathbf{x}})+\mathbf{A}^{H}\mathrm{Diag}(\mathbf{1}\oslash\mathbf{v}_{2\mathbf{z}})\mathbf{A} \right)}^{-1}, \label{eq33a}\\
     \hat{\mathbf{x}}_{2}&=\mathbf{Q}_{2\mathbf{x}} \left(\mathbf{r}_{2\mathbf{x}}\oslash\mathbf{v}_{2\mathbf{x}}+\mathbf{A}^{H}\mathbf{r}_{2\mathbf{z}}\oslash\mathbf{v}_{2\mathbf{z}}\right), \label{eq33b}\\
     \mathbf{Q}_{2\mathbf{z}}&=\mathbf{A}\mathbf{Q}_{2\mathbf{x}}\mathbf{A}^{H}, \label{eq33c} \\
    \hat{\mathbf{z}}_{2}&=\mathbf{A}\hat{\mathbf{x}}_{2}. \label{eq33d}
    \end{align}
\end{subequations}
 Compute the extrinsic information of $\mathbf{z}$ %from the linear space
  \begin{subequations}\label{eq34}
    \begin{align}
     % \boldsymbol{\eta}_{2\mathbf{z}}&=\mathbf{1}\oslash\mathbf{d}(\mathbf{Q}_{2\mathbf{z}}), \label{eq34a} \\
     \mathbf{v}_{1\mathbf{z}}&= \mathbf{1}\oslash {\left(\mathbf{1}\oslash\mathbf{d}(\mathbf{Q}_{2\mathbf{z}}) - \mathbf{1}\oslash\mathbf{v}_{2\mathbf{z}} \right)}, \label{eq34b} \\
     \mathbf{r}_{1\mathbf{z}}&=\mathbf{v}_{1\mathbf{z}} \odot {\left( \hat{\mathbf{z}}_{2}\oslash\mathbf{d}(\mathbf{Q}_{2\mathbf{z}}) - \mathbf{r}_{2\mathbf{z}} \oslash \mathbf{v}_{2\mathbf{z}} \right)}. \label{eq34c}
    \end{align}
\end{subequations}
\end{enumerate}
 \EndWhile
\State \Return the recovered signal $\hat{\mathbf{x}}_{1}$.
\end{algorithmic}
\end{footnotesize}
\end{algorithm}

Algorithm 1 specifies the iterative procedure of the GEC-SR.
In Algorithm 1, the posterior mean and the variance of $\mathbf{z}$ and $\mathbf{x}$ are obtained from (\ref{eq27}) and (\ref{eq31}), respectively.
We take the expectation and variance in (\ref{eq31a}) and (\ref{eq31b}) with respect to the posterior probability
 \begin{equation}\label{eq15}
    p_{1}(\mathbf{x}|\mathbf{r}_{1\mathbf{x}},\mathbf{v}_{1\mathbf{x}})=\frac{e^{\log p(\mathbf{x})-\|\mathbf{x}-\mathbf{r}_{1\mathbf{x}}\|_{\mathbf{v}_{1\mathbf{x}}}^{2}}}{\int e^{\log p(\mathbf{x})-\|\mathbf{x}-\mathbf{r}_{1\mathbf{x}}\|_{\mathbf{v}_{1\mathbf{x}}}^{2}} d\mathbf{x}},
 \end{equation}
 where
\begin{equation}\label{eq16}
   \|\mathbf{a}\|_{\mathbf{v}}^{2} \triangleq \sum_{n=1}^{N} \frac{|a_{n}|^2}{v_{n}}.
\end{equation}

We can calculate the expectation and variance on each entry of $\mathbf{x}$ separately because the prior $p(\mathbf{x})$ is separable, and thus
we omit index $n$ in the following expressions.
Using the Gaussian reproduction property \cite{GPML}, we can obtain the explicit componentwise expression
\begin{align}
    \mathtt{E}\{x | r, v\} &= C
    \frac{r\rho^{-1}}{v+\rho^{-1}}, \label{eq17} \\
  \mathtt{Var}\{x | r, v \} &=C{\left(\frac{v\rho^{-1}}{v+\rho^{-1}}+\left|\frac{r\rho^{-1}}{v+\rho^{-1}}\right|^{2} \right)}-|\hat{x}|^{2}, \label{eq18}
\end{align}
where
\begin{equation}\label{eq19}
 C=\frac{\rho\mathcal{N}_{\mathbb{C}}(0;r,v+\rho^{-1})} {(1-\rho)\mathcal{N}_{\mathbb{C}}(0;r,v)+\rho\mathcal{N}_{\mathbb{C}}(0;r,v+\rho^{-1})}.
\end{equation}

Similarly, the posterior mean and variance of $\mathbf{z}$ in (\ref{eq27a}) and (\ref{eq27b}) are taken with respect to the posterior
 \begin{equation}\label{eq20}
    p_{1}(\mathbf{z}|\mathbf{r}_{1\mathbf{z}},\mathbf{v}_{1\mathbf{z}})=\frac{e^{\log p(\mathbf{y}|\mathbf{z})-\|\mathbf{z}-\mathbf{r}_{1\mathbf{z}}\|_{\mathbf{v}_{1\mathbf{z}}}^{2}}}{\int e^{\log p(\mathbf{y}|\mathbf{z}) -\|\mathbf{z}-\mathbf{r}_{1\mathbf{z}}\|_{\mathbf{v}_{1\mathbf{z}}}^{2}} d\mathbf{z}}.
 \end{equation}
The mean and variance can also be computed in a componentwise manner.
(\ref{eq27a}) and (\ref{eq27b}) are nonlinear because of the quantization, and their explicit expressions are provided in \cite{wenJCD}.

Under the linear constraint $\mathbf{z}=\mathbf{A}\mathbf{x}$, the estimation of the posterior mean and covariance matrix of $\mathbf{x}$ are obtained in (\ref{eq29b}) and (\ref{eq29a}) with the corresponding posterior probability
\begin{equation}\label{eq24}
 p_{2}(\mathbf{x}|\mathbf{r}_{2},\mathbf{v}_{2})=\frac{e^{-\|\mathbf{x}-\mathbf{r}_{2\mathbf{x}}\|_{\mathbf{v}_{2\mathbf{x}}}^{2}-\|\mathbf{z}-\mathbf{r}_{2\mathbf{z}}\|_{\mathbf{v}_{2\mathbf{z}}}^{2}} }
 {\int e^{-\|\mathbf{x}-\mathbf{r}_{2\mathbf{x}}\|_{\mathbf{v}_{2\mathbf{x}}}^{2}-\|\mathbf{\mathbf{z}}-\mathbf{r}_{2\mathbf{z}}\|_{\mathbf{v}_{2\mathbf{z}}}^{2}} d\mathbf{x}}.
\end{equation}
The posterior mean and covariance matrix of $\mathbf{z}$ can be obtained in (\ref{eq33}) following the linear space of $\mathbf{z}=\mathbf{A}\mathbf{x}$.

\subsection{Relation of GEC-SR and Initial GEC}
In the introduction, we mention that our work is inspired by the ``initial GEC'' algorithm from \cite{GEC}, which considers the standard linear measurement and GLM. However, our algorithm is different from the initial GEC in terms of the update manner. In the GEC-SR, we first estimate $\mathbf{z}$ from the nonlinear measurements $\tilde{\mathbf{y}}$ followed by estimating the signal $\mathbf{x}$ using the prior information from module C, whereas the initial GEC estimates $\mathbf{x}$ and $\mathbf{z}$ simultaneously. In addition, before computing the mean and covariance of $\mathbf{z}$ in (\ref{eq33c}) and (\ref{eq33d}), we compute the mean and covariance of $\mathbf{x}$ once again in (\ref{eq33a}) and (\ref{eq33b}). Because of these modifications, the GEC-SR algorithm converges faster than initial GEC and can agree perfectly with the theoretical SE analysis that predicted by the replica method. We will show the theoretical SE analysis in the next section.

\subsection{Relation of GEC-SR and GTurbo-SR}
GTurbo-SR \cite{LiuT} is a promising algorithm to recover sparse signals from nonlinear measurements, and the idea uses the turbo principle in iterative decoding to compute the extrinsic messages of $\mathbf{x}$ and $\mathbf{z}$. A visual examination of the GEC-SR shows many similarities with the GTurbo-SR in terms of the iterative approach. In particular, the posterior probabilities of $\mathbf{x}$ and $\mathbf{z}$ in the GEC-SR are identical to those in the GTurbo-SR. Similarly, the computation of extrinsic information in the GEC-SR is also identical to the one in the GTurbo-SR. However, GTurbo-SR only considers the sensing matrix $\mathbf{A}$ as a partial DFT matrix, while general matrices can be applied in the GEC-SR.
 If we replace $\mathbf{A}$ by a partial DFT matrix in the GEC-SR, the GEC-SR is reduced to the GTurbo-SR.

\section{State Evolution}
In this section, we show the SE equations of the GEC-SR.
From the statistical
mechanics perspective, the iterative procedure of the GEC-SR is equivalent to finding the saddle points of the free energy defined by
\begin{equation}
    {\cal F} = - \frac{1}{N} \mathtt{E} \{ \log p(\tilde{\mathbf{y}}) \}.
\end{equation}
The calculation of ${\cal F}$ is very difficult. Fortunately, the replica method from statistical physics provides a highly sophisticated procedure to address this calculation.
In the calculation, we use the assumptions that $N,M\rightarrow\infty$ while keeping $M/N=\alpha$ fixed and finite.
Only the final analytical results in Proposition 1 are shown because of space limitation.

Proposition 1 involves several new parameters.
%First, we characterize the sensing matrix $\mathbf{A}$ by its eigenvalue spectrum
%\begin{equation}\label{eq33}
%  \mathbf{A}\mathbf{A}^{H}=\mathbf{V}^{H}\boldsymbol{\Lambda}\mathbf{V},
%\end{equation}
%where $\mathbf{V}$ is the $N \times N$ unitary matrix and $\boldsymbol{\Lambda}$ is a diagonal matrix whose diagonal elements are the corresponding eigenvalues $\{\lambda_{i},i=1,\ldots,N\}$.
Most parameters (except for some auxiliary parameters) can be illustrated systematically by a scalar channel
\begin{equation}\label{eq34}
  r=x+w,
\end{equation}
where $w\sim\mathcal{N}_{\mathbb{C}}(w;0,\eta^{-1})$. The MMSE estimate of (\ref{eq34}) is given by
\begin{equation}\label{eq35}
  \mathtt{E}\{x|r\}=\int xp(x|r)dx,
\end{equation}
where $p(x|r)=\frac{p(r|x)p(x)}{p(r)}$ and $p(r|x)=\frac{\eta}{\pi}e^{-\eta|r-x|^{2}}$.
We define the MMSE of this estimator as
\begin{equation}\label{eq36}
  \mathrm{mmse}(\eta)=\mathtt{E} {\left\{|x-\mathtt{E}\{x|r\}|^{2}\right\}},
\end{equation}
where the expectation is taken over the joint distribution $p(r,x)=p(r|x)p(x)$. If $x$ follows the Bernoulli-Gaussian distribution (\ref{eq2}),
$\mathrm{mmse}(\eta)$ can be obtained explicitly \cite{SE}
\begin{multline}
\mathrm{mmse}(\eta) = 1-\frac{\eta}{\eta\rho^{-1}+1} \\
     {\times}\int Dz_{c}\frac{|z|^{2}}{\rho+(1-\rho)e^{-|z|^{2}\eta\rho^{-1}}(\eta\rho^{-1}+1)}.
\end{multline}
For ease of expressions, we define two auxiliary equations:
\begin{align}
  \tilde{\eta}_{z}&= \mathtt{E} \Bigg\{\frac{\frac{1} {v_{z}}\frac{1} {v_{x}}}{\frac {1}{v_{x}} + \frac{\lambda}{v_{z}}} \Bigg\} , \label{eq37} \\
  P_{x}-\tilde{\eta}_{x}&= (1-\alpha)v_{x} + \alpha \mathtt{E} \Bigg\{ \frac{1}{\frac{1}{v_{x}} + \frac{\lambda} {v_{z}}} \Bigg\} , \label{eq38}
\end{align}
where $\lambda$ is the eigenvalues of $\mathbf{A}\mathbf{A}^{H}$, the expectation with respect to $\lambda$ is defined by $ \mathtt{E}\{ f(\lambda)\} = \frac{1}{M} \sum_{i=1}^{M}f(\lambda_i) $, $(\tilde{\eta}_{x}, \tilde{\eta}_{z},v_{x},v_{z})$ will be given in Proposition 1, and
$(P_{x},P_{z})$ have been defined in (\ref{eq:defPxPz}).
In addition, we denote $\Psi'(\tilde{y};z,c^2) = \frac{\partial \Psi(\tilde{y};z,c^2)}{\partial z}$.

\begin{Proposition}
The saddle points of the free energy can be obtained by
\begin{align*}
  & \mbox{Initial $t=0$, $v_{x}^{0}= P_{x}$ and $\eta_{z}^{0}=0$.}   \\
  & t = 0, 1,2, \ldots\\
  &1)~ \tilde{\eta}_{z}^{t}:=\sum_{\tilde{y} \in \mathcal{R}_{B}}\int Dz\frac{\bigg(\Psi'\left(\tilde{y};\sqrt{\frac{\eta_{z}^{t}}{2}}z, \frac{\sigma^2 +P_{z}-\eta_{z}^{t}}{2} \right)\bigg)^{2}}{\Psi\left(\tilde{y};\sqrt{ \frac{\eta_{z}^{t}}{2}}z, \frac{\sigma^2 +P_{z}-\eta_{z}^{t}}{2} \right)};   \\
  &\hspace{0.45cm} v_{z}^{t+1}:= \frac{1}{\tilde{\eta}_{z}^{t}} - {(P_{z}- \eta_{z}^{t})}; \\
  &2)~ \mbox{Get $P_{x}-\tilde{\eta}_{x}^{t}$ using (\ref{eq38}) for a given $(v_{x}^{t},v_{z}^{t+1})$,}   \\
  &\hspace{0.45cm} \eta_{x}^{t+1}:= \frac{1}{P_{x}-\tilde{\eta}_{x}^{t}} - \frac{1}{v_{x}^{t}};  \\
  &3)~ v_{x}^{t+1}:= \left( \frac{1}{\mathrm{mmse}(\eta_{x}^{t+1})}-\eta_{x}^{t+1} \right)^{-1} ;   \\
  &4)~ \mbox{Get $\tilde{\eta}_{z}^{t+1}$ using (\ref{eq37}) for a given $(v_{x}^{t+1},v_{z}^{t+1})$;}   \\
  &\hspace{0.45cm} P_{z}-\eta_{z}^{t+1}:=\frac{1}{\tilde{\eta}_{z}^{t+1}}-  {v_{z}^{t+1}}.
\end{align*}
\hspace{8.5cm} $\blacksquare$
\end{Proposition}

As $ t\rightarrow \infty$, $\{ \eta_{x}^{t}, \eta_{z}^{t} \}$ converges to a saddle point of the free energy.
The above iterative expressions also correspond to the SEs of the GEC-SR in Algorithm 1. In particular, $\mathrm{mmse}(\eta_{x}^{t})$ represents the MSE of $\hat{\mathbf{x}}$.

If $\mathbf{A}$ is obtained by the random selection of a set of rows from the standard DFT matrix, then $\mathbf{A}$ is the row-orthogonal matrix with eigenvalues $\lambda_{i}=1$ for $i=1,\ldots,N$. By combining all the coupled equations, we finally obtain
\begin{align}
    &\tilde{\eta}_{z}^{t}:=\sum_{\tilde{y} \in \mathcal{R}_{B}} \int Dz \frac{\bigg(\Psi'\left(\tilde{y}; \sqrt{ \frac{P_{z}-v_{x}^{t}}{2} } z, \frac{\sigma^2 +v_{x}^{t}}{2} \right)\bigg)^{2}}{\Psi\left(\tilde{y}; \sqrt{ \frac{P_{z}-v_{x}^{t}}{2} } z, \frac{\sigma^2 +v_{x}^{t}}{2} \right)}, \label{eq44} \\
  &\eta_{x}^{t+1}:=\bigg(\frac{1}{\alpha \tilde{\eta}_{z}^{t}}-v_{x}^{t}  \bigg)^{-1}, \label{eq45} \\
  &v_{x}^{t+1}:=\bigg(\frac{1}{\mathrm{mmse}(\eta_{x}^{t+1})}-\eta_{x}^{t+1} \bigg)^{-1}. \label{eq46}
\end{align}
The above iterative equations agree with those in the GTurbo-SR \cite{LiuT}.

\section{Numerical results}
In this section, we conduct numerical experiments to verify the accuracy
of our analytical results.
In all the cases, we consider the recovery $\mathbf{x}$ from the quantized output $\tilde{\mathbf{y}}$ constructed from (\ref{eq10}), where $\mathbf{x}$ is drawn i.i.d., zero-mean Bernoulli-Gaussian with $\rho=0.4$. The noise level $\sigma^{2}$ is set as $10^{-5}$.
%For comparison, the simulation scenarios completely follow those presented in \cite{JMa,LiuT}, where the system parameters are set as follows: $\alpha=0.7$, $N=8192$, $M=5734$, and ${\rm SNR}=50$dB. The metric MSE is defined as
% {\rl Because it is difficult to do experiments with a big system, the system parameters are set as follows: $\alpha=0.7$, $N=819$, $M=573$, and ${\rm SNR}=50$dB.}
The metric MSE is defined as
\begin{equation}\label{eq21}
 {\rm MSE} = \frac{\|\mathbf{x}-\hat{\mathbf{x}}_{1}\|^{2}}{\|\mathbf{x}\|^{2}}=\frac{\|\mathbf{x}-\hat{\mathbf{x}}_{1}\|^{2}}{N}.
\end{equation}
We use the typical uniform quantizer with quantization step size $\Delta=2^{1-B}$, where $B$ is the quantization resolution. The simulation results are obtained by averaging over $2,000$ realizations.
\begin{figure}
  \centering
  \includegraphics[width=3.5in]{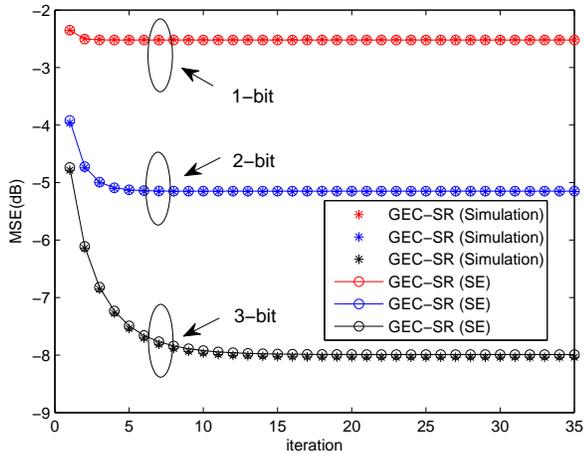}
  \caption{Simulated and analytical MSEs of the GEC-SR under different quantization levels. The singular values of sensing matrix $\mathbf{A}\in {\mathbb C}^{5734 \times 8192}$ are set as $[\lambda_1\mathbf{1}_{M_{1}} ~\lambda_2\mathbf{1}_{M_{2}}]$ with $M_{1}=5000$, $M_{2}=734$, and $(\lambda_1,\lambda_2)=(1,3)$. }\label{fig.sim.GEC}
\end{figure}
\begin{figure}
  \centering
  \includegraphics[width=3.5in]{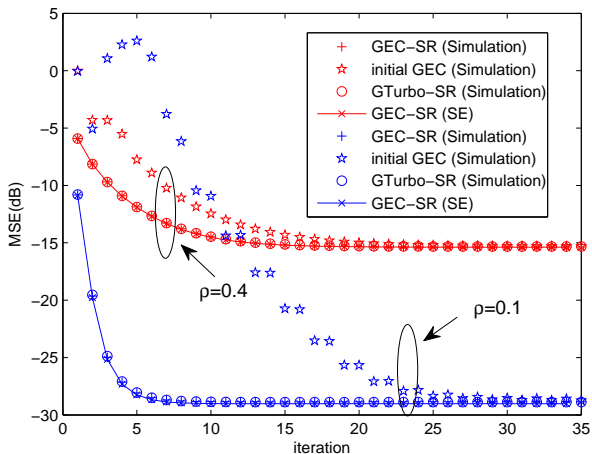}
  \caption{MSE results of Algorithm 1, GTurbo-SR, and initial GEC with partial DFT sensing
matrix under different sparasity levels.}\label{fig.sim.se}
\end{figure}

Figure \ref{fig.sim.GEC} plots the average MSEs achieved by the GEC-SR and the theoretical result derived by the replica method under a general matrix. We constructed $\mathbf{A}\in {\mathbb C}^{5734 \times 8192}$ from the singular value decomposition $\mathbf{A}=\mathbf{U}\mathbf{D}\mathbf{V}^{T}$, where unitary matrices $\mathbf{U}$ and $\mathbf{V}$ are drawn uniformly with respect to the Haar measure. The singular values are set as $[\lambda_1\mathbf{1}_{M_{1}} ~\lambda_2\mathbf{1}_{M_{2}}]$ with $M_{1}=5000$, $M_{2}=734$, and $(\lambda_1,\lambda_2)=(1,3)$.
%Figure \ref{fig.sim456.GEC} shows the corresponding results for another experiment under a larger system, where $N=8192$, $M=5734$, and the singular values are a geometric series with $\lambda_{1}=0.01$ and $\lambda_{n}/ \lambda_{n-1}=1.0005$.
%%Particularly, $\mathbf{A}$ is rotationally invariant.
%The figures illustrate that the GEC-SR shows excellent agreement with
%the theoretical SE expressions shown in Proposition 1 for a general matrix.

Figure \ref{fig.sim.se} shows the corresponding MSEs of Algorithm 1, GTurbo-SR \cite{LiuT}, and initial GEC \cite{GEC} with partial DFT sensing matrix under different sparasity levels. The quantization level is $3\text{-}\mathrm{bit}$. For comparison, the simulation scenarios completely follow those presented in \cite{JMa,LiuT}, where the system parameters are set as follows: $\alpha=0.7$, $N=8192$, and $M=5734$. The figure clearly demonstrates that the GEC-SR is identical to the GTurbo-SR when partial DFT is considered, and the SE analysis precisely predicts the per iteration performance. In addition, the initial GEC cannot coverage to the fixed-point when the signal is very sparse, but our GEC-SR algorithm is more robust because of the different update manner.

\section{Conclusion}
In this paper, we developed a computationally feasible signal recovery approximation scheme called GEC-SR for nonlinear measurements affected by quantization.
We showed that the performance of the GEC-SR is superior to initial GEC for general sensing matrices, and the GEC-SR is reduced to GTurbo-SR for partial DFT sensing matrices. Finally, we presented the SE analysis to precisely describe the asymptotic behavior of the GEC-SR algorithm.

\end{document}